\documentclass[useAMS,usenatbib]{mn2e}

\usepackage[latin1]{inputenc}
\usepackage{graphicx}
\usepackage{epstopdf}

\usepackage{subfigure}
\usepackage{rotate}
\usepackage{amsmath, float}
\usepackage{txfonts}
\usepackage{fleqn}
\usepackage{color}
\usepackage{comment}

\usepackage{mathrsfs}  
\usepackage{booktabs}
\usepackage{colortbl} 

\newcolumntype{G}{>{\columncolor[gray]{0.8}}l} 

\newcommand{\be}{\begin{equation}}
\newcommand{\ee}{\end{equation}}
\newcommand{\bdm}{\begin{displaymath}}
\newcommand{\edm}{\end{displaymath}}
\newcommand{\bea}{\begin{multline}}
\newcommand{\eea}{\end{multline}}
\newcommand{\ba}{\begin{align}}
\newcommand{\ea}{\end{align}}
\newcommand{\btZ}{\beta_z}

\newcommand\al{\alpha}
\newcommand\bt{\beta}
\newcommand\dl{\delta}
\newcommand\tht{\phi}
\newcommand\sg{\sigma}
\newcommand\inc{i}

\def\simlt{\mathrel{\hbox{\rlap{\hbox{\lower4pt\hbox{$\sim$}}}\hbox{$<$}}}}
\def\simgt{\mathrel{\hbox{\rlap{\hbox{\lower4pt\hbox{$\sim$}}}\hbox{$>$}}}}

\title[X-rays from PWNe and magnetic turbulence]
{Modeling the effect of small-scale magnetic turbulence on the X-ray
  properties of Pulsar Wind Nebulae}
\author[N. Bucciantini, R. Bandiera, B. Olmi]{
N. Bucciantini$^{1,2,3}$\thanks{E-mail: niccolo@arcetri.astro.it},
R. Bandiera$^{1}$, B. Olmi$^{2,1,3}$, L. Del Zanna$^{2,1,3}$ \\
$^{1}$INAF - Osservatorio Astrofisico di Arcetri, Largo E. Fermi 5,
I-50125 Firenze, Italy\\
$^{2}$Dipartimento di Fisica e Astronomia, Universit\`a degli Studi di Firenze, Via G. Sansone 1, 
I-50019 Sesto F.~no  (Firenze), Italy\\
$^{3}$INFN - Sezione di Firenze, Via G. Sansone 1, I-50019 Sesto F.~no  (Firenze), Italy}

\begin{document}
 
\date{Accepted / Received}

\maketitle

\label{firstpage}

\begin{abstract}
Pulsar Wind Nebulae (PWNe) constitute an ideal astrophysical environment to
test our current understanding of relativistic plasma processes. It is
well known that magnetic fields play a crucial role in their dynamics
and emission properties. At present, one of the main issues concerns the
level of magnetic turbulence present in these systems, which in the
absence of  space resolved X-ray polarization measures cannot be
directly constrained.  In this work we investigate, for the first time using simulated synchrotron maps, the effect of a small scale fluctuating component of the magnetic
field  on the emission properties in X-ray. We illustrate how to include the effects of
a turbulent component in standard emission models for PWNe,
and which consequences are expected in terms of net emissivity and
depolarization, showing that the X-ray surface brightness maps can
provide already some rough constraints. We then apply our analysis to the Crab and Vela
nebulae and, by comparing our model with Chandra and Vela data, we
found that the typical energies in the  turbulent component of the
magnetic field are about 1.5 to 3 times the one in the ordered field.

\end{abstract}


\section{Introduction}

Pulsar Wind Nebulae (PWNe) are bubbles of relativistic particles and
magnetic field arising from the interaction of the relativistic pulsar
wind with the ambient medium (ISM or supernova remnant). They shine in
non-thermal (synchrotron and inverse Compton) radiation
 in a broad range of frequencies from radio wavelengths to
$\gamma$-rays (see \citealt{Gaensler_Slane06a} for a review). At
X-rays, MANY PWNe show an axisymmetric feature known as
 {\it jet-torus structure}. This feature has been observed by now in a
number of PWNe, among which the Crab nebula \citep{Weisskopf_Hester+00a},
Vela \citep{Helfand_Gotthelf+01a,Pavlov_Kargaltsev+01a} and MSH 15-52 \citep{Gaensler_Arons+02a,DeLaney_Gaensler+06a}, to name just a few. It is now commonly accepted that this
structure arises due to the interplay of the anisotropic energy flux
in the wind with the toroidal magnetic field, as confirmed by a long
series of numerical simulations \citep{Komissarov_Lyubarsky03a,Komissarov_Lyubarsky04a,Del-Zanna_Amato+04a,Del-Zanna_Volpi+06a,Volpi_Del-Zanna+07a,Camus_Komissarov+09a,Porth_Komissarov+14a,Olmi_Del-Zanna+14a,Olmi_Del-Zanna+16a}. In general, two dimensional models are built on the axisymmetric
assumption of a purely toroidal magnetic field, while three dimensional
models have usually much lower resolution, and can only investigate
large scale deviations from axisymmetry. However, recently, several
arguments have been put forward, advocating the presence of small
scale turbulence in PWNe: the presence of a large diffuse X-ray halo
at distances in excess of the naive expectation for synchrotron
cooling and advection \citep{Tang_Chevalier12a,Buhler_Blandford14a,Zrake_Arons16a}; the suggestion that radio emitting
particles could be continuously reaccelerated in the body of the
nebula  \citep{Olmi_Del-Zanna+15a,Tanaka_Asano16a}; the observation or recurrent $\gamma$-ray flares,
requiring localized strong current sheets \citep{Uzdensky_Cerutti+11a}.

Radio polarization maps are available, but, being radio emission
dominated by the outer region of PWNe, where the effects of the
interaction with the SNR are stronger, they provide at best a good estimate of the
degree of ordered versus disordered magnetic field for the overall
nebula, but cannot be used to constrain the conditions in the region close to the
termination shock, where most of the variability and the acceleration
processes take place. In the Crab nebula, the polarized fraction in radio is on average
$\sim 16\%$ \citep{Conway71a,Ferguson73a,Velusamy85a,Aumont_Conversi+10a} with
peaks up to $30\%$, lower than in optical, where the average polarized
fraction is $\sim 25\%$ \citep{Velusamy85a}. Moreover the polarized
flux in radio anti-correlates with the location of the X-ray torus. The values of the
radio polarization are consistent with a largely turbulent magnetic
field in the outer part of the nebula (Bucciantini et al. in preparation).

Recently  high resolution HST observations in polarized light have been presented
both for the inner region of the Crab nebula \citep{Moran_Shearer+13b}, and for the
Vela PWN \citep{Moran_Mignani+14a}. The study of the Crab nebula focused on the brightest
optical features, namely the {\it
  knot} and the {\it wisps}, which are shown to have typical polarized
fraction of about 60\% and 40\% respectively. The results in
the  Crab nebula are
consistent  with the general picture of a strongly ordered toroidal
magnetic field just downstream of the termination shock, with a
possible hint of the development of turbulence: the polarized fraction
of the wisps is lower than the one in the knot, and numerical
simulations suggest the former to be slightly more downstream than
the latter. Vela is not detected in polarized optical light, but this might not be
very constraining given that there is no optical counterpart observed
for this nebula \citep{Marubini_Sefako+15a}. The major problem with optical emission is
that there is usually a large foreground, often polarized (see for
example the polarization analysis of the Crab nebula presented in
\citealt{Hester08a}), and the jet-torus structure, which is so prominent in X-ray, is much fainter. 
Ideally one would like to probe these systems using X-ray
polarimetry \citep{Bucciantini10a}, and there is a large interest in the scientific
community for such objective \citep{Soffitta_Barcons+13a,Weisskopf_Ramsey+16a}. Incidentally, the Crab nebula
is at the moment the only object with a polarization detected in X-ray
\citep{Weisskopf_Silver+78a}. 

In the past years several models have been presented to simulate the
X-ray emission map of PWNe (with a particular focus on Crab): ranging
from simplified toy-models \citep{Ng_Romani04a,Schweizer_Bucciantini+13a}, to more complex multi-dimensional
time-dependent simulations \citep{Volpi_Del-Zanna+07a,Porth_Komissarov+14a,Olmi_Del-Zanna+16a}. Starting from the early work of
\citet{Bucciantini_del-Zanna+05a}, polarization have also been modelled using in general a
magnetic field geometry derived from numerical simulations
\citep{Del-Zanna_Volpi+06a,Porth_Komissarov+14a}. However the presence of small scale turbulence and its
effects both on the polarized fraction and on the emissivity pattern,
have never been taken into account before. 

Here we present synthetic polarization maps of PWNe, taking into
account the presence of small scale magnetic turbulence at a subgrid
level. In Sect.~2 we illustrate how the recipe for total and
polarized emission can be corrected to take into account magnetic
turbulence. In Sect.~3, using a simple toy model, we show how the
effects of turbulence manifest in the intensity map, and are thus in
principle already accessible from X-ray imaging. In Sect.~4 we
present a semi-analytical model for a thin-ring that allows us to
derive simple formulae showing the typical degeneracy between Doppler
boosting and turbulence. In Sect.~5 we
apply our results to the Crab and Vela PWNe, objects that have been considered as primary
targets for future X-ray polarimetric observations \citep{Weisskopf_Ramsey+16a}.

\section{Polarization recipes and subgrid model}

Let us begin by recalling the general recipe to compute the
synchrotron intensity and polarization properties, taking into account
relativistic Doppler boosting effects, in the case of a fully ordered
(at least on the scale of the fluid element under consideration)
magnetic field. A complete derivation can be found in
\citet{Del-Zanna_Volpi+06a}. For electrons having a power-law
distribution
\begin{equation}
n(\epsilon)= K \epsilon^{-(2\alpha+1)},
\label{eq:fdist}
\end{equation}
where $\epsilon$ is the energy in units $m_e c^2$, and belonging to a fluid element with comoving magnetic field
$\boldsymbol{B}'$, and velocity with respect to the observer
$\bf{v}=\boldsymbol{\beta}$$c$ (corresponding to a Lorentz
factor $\gamma$), the emissivity toward the
observer at a frequency $\nu$ is:
\begin{equation}
j(\nu,\boldsymbol{n}) = C\mid\boldsymbol{B}'\times \boldsymbol{n}'
\mid^{\alpha+1} D^{\alpha+2} \nu^{-\alpha}\,,
\label{eq:j}
\end{equation}
where $C$ is given by synchrotron theory
\begin{equation}
C=\frac{\sqrt{3}}{4}\frac{\alpha+5/3}{\alpha+1}\Gamma\left(\frac{\alpha
  +5/3}{2}\right)\Gamma\left(\frac{\alpha
  +1/3}{2}\right)\frac{e^3}{mc^2}\left(\frac{3e}{2\upi m c}
\right)^\alpha K\,,
\end{equation}
$\boldsymbol{n}'$ is the direction of the observer measured in the
comoving frame, related to the one measured in the observer frame
$\boldsymbol{n}$ by
\begin{equation}
\boldsymbol{n}'= D\left[\boldsymbol{n}+\left(\frac{\gamma^2}{\gamma+1}\boldsymbol{\beta}\cdot\boldsymbol{n}-\gamma  \right)\boldsymbol{\beta}   \right]\,,
\end{equation}
$D$ is the Doppler boosting factor
\begin{equation}
D=\frac{1}{\gamma(1-\boldsymbol{\beta}\cdot\boldsymbol{n})}\,,
\label{eq:doppler}
\end{equation}
and the comoving magnetic field can be computed from the one measured
in the observer frame as
\begin{equation}
\boldsymbol{B}'=\frac{1}{\gamma}\left[\boldsymbol{B}+\frac{\gamma^2}{\gamma+1}(\boldsymbol{\beta}\cdot\boldsymbol{B})\boldsymbol{\beta} \right]\,,
\label{eq:bcomov}
\end{equation}
giving
\begin{equation}
\mid\boldsymbol{B}'\times
\boldsymbol{n}'\mid=\frac{1}{\gamma}\sqrt{B^2-D^2(\boldsymbol{B}\cdot\boldsymbol{n})^2+2\gamma
D (\boldsymbol{B}\cdot\boldsymbol{n})(\boldsymbol{B}\cdot\boldsymbol{\beta})}.\,\label{eq:bcrossn}
\end{equation}
Let us consider a Cartesian observer's reference frame in which $X$
lies along the line of sight $\boldsymbol{n}$ and $Y$ and $Z$ are in the
plane of the sky. At this point it is possible to compute the maps of
the various Stokes parameters
integrating the contribution of each fluid element along the line of sight through the nebula, according to
\begin{eqnarray}
I(\nu,Y,Z)&=&\int_{-\infty}^{\infty}j(\nu,X,Y,Z)\,{\rm d}X\,, \\
Q(\nu,Y,Z)&=&\frac{\alpha+1}{\alpha+5/3}\int_{-\infty}^{\infty}j(\nu,X,Y,Z)\cos{2\chi}\,{\rm d}X\,, \\
U(\nu,Y,Z)&=&\frac{\alpha+1}{\alpha+5/3}\int_{-\infty}^{\infty}j(\nu,X,Y,Z)\sin{2\chi}\,{\rm d}X\, ,
\end{eqnarray}
where  the local polarization position angle $\chi$ is the angle of
the emitted electric field vector $\boldsymbol{e}$ in the plane of the
sky. This electric field is related to the one measured at emission in
the comoving frame $\boldsymbol{e}'$ by
\begin{equation}
\boldsymbol{e}=\gamma\left[\boldsymbol{e}'-\frac{\gamma}{\gamma+1}(\boldsymbol{\beta}\cdot\boldsymbol{e}')\boldsymbol{\beta}-\boldsymbol{\beta}\times(\boldsymbol{n}'\times\boldsymbol{e}') \right]\,.
\end{equation}
In Ideal MHD it is possible to introduce an auxiliary vector $
\boldsymbol{q}$ defined as
\begin{eqnarray}
q_Y=(1-\beta_X)B_Y+\beta_Y B_X\,,\quad \quad q_Z=(1-\beta_X)B_Z+\beta_ZB_X\,, \label{eq:stokesang1}
\end{eqnarray}
such that
\begin{eqnarray}
\cos{2\chi} = \frac{q_Y^2-q_Z^2}{q_Y^2+q_Z^2}\,,\quad\quad\sin{2\chi}=-\frac{2q_Yq_Z}{q_Y^2+q_Z^2}\,.
\label{eq:stokesang2}
\end{eqnarray}
In a recent paper \citet{Bandiera_Petruk16a} have shown that the
effect of the small scale magnetic field fluctuations on the total and
the polarized emissivity can be computed analytically,  considering a
fluid element with a net average field and assuming that the small
scale fluctuations can be described by an isotropic random Gaussian
field with variance $(B' \sigma)^2$ in each direction. The emission is computed
considering the electrons to be distributed in the nebula with a
power-law distribution function, as specified in Eq.~\ref{eq:fdist}. The
variance is just a measure of the ratio of the energy  $\delta E$ in the small scale
fluctuating components over the energy $E$ in the ordered
component of the comoving magnetic field: $\delta E/E=3\sigma^2$. These small
scale fluctuations contribute to the total emissivity, which rises
linearly with the energy in the fluctuating components (as long as
this energy is smaller than the one associated to the net average
magnetic field). They however contribute much less to the polarized
intensity, because of the assumption that they are randomly
distributed in orientation. The net effect is to reduce the polarized
fraction. Interestingly the depolarization is almost insensitive to
the value of $\alpha$.  One can introduce two correction coefficients
defined as 
\begin{align}
\xi(\alpha,B_\perp,\sigma)&\!=\!\Gamma\left(\frac{3+\alpha}{2}\right){_1}F_1\left(-\frac{1+\alpha}{2},1,-\frac{B_\perp^2}{2B'^2\sigma^2}
  \right)\left( \frac{B_\perp}{\sqrt{2}B'\sigma} \right)^{-(1+\alpha)}\\
\zeta(\alpha,B_\perp,\sigma)&\!=\!\frac{1}{2}\Gamma\left(\frac{5+\alpha}{2}\right){_1}F_1\left(\frac{1-\alpha}{2},3,-\frac{B_\perp^2}{2B'^2\sigma^2}
  \right)\left( \frac{B_\perp}{\sqrt{2}B'\sigma} \right)^{(1-\alpha)}
\end{align}
where $_1F_1(a,b,x)$  is the Kummer confluent hypergeometric function,
and  $B_\perp = \mid \boldsymbol{B}'\times\boldsymbol{n}' \mid$ is the component
of the average comoving magnetic
field perpendicular to the direction of emission,  that can be taken from a large scale
simulation, or a toy model, leaving only $\sigma$ as a free
parameter.

Then one can
compute maps corrected for small scale fluctuations as
\begin{eqnarray}
I(\nu,Y,Z) \!&=&\!\!\int_{-\infty}^{\infty}\xi(\alpha,B'_\perp,\sigma)j(\nu,X,Y,Z)\,{\rm d}X\\
Q(\nu,Y,Z) \!&=&\!\!\frac{\alpha+1}{\alpha+5/3}\int_{-\infty}^{\infty}\zeta(\alpha,B'_\perp,\sigma)j(\nu,X,Y,Z)\cos{2\chi}\,{\rm d}X\\
U(\nu,Y,Z) \!&=&\!\!\frac{\alpha+1}{\alpha+5/3}\int_{-\infty}^{\infty}\zeta(\alpha,B'_\perp,\sigma)j(\nu,X,Y,Z)\sin{2\chi}\,{\rm d}X.
\end{eqnarray}

It can  be shown
that it is not correct to model a fluctuating component just by adding
a constant unpolarized emission on top of the one derived assuming a
totally ordered field. A cautionary remark
is here in order: the corrections derived by  \citet{Bandiera_Petruk16a}, are formally
valid only in the limit $\beta\rightarrow 0$, when the isotropic
assumption in the comoving frame corresponds to the isotropic
assumption in the observer's frame. In case of strongly relativistic
motions this is in general not  true. The correction coefficients in
such a case are defined properly in the comoving frame, and one would
need to perform a Lorentz transformation of the polarization tensor to get
the correct result in the observer's reference frame. However the typical bulk flow
in the body of PWNe (and in particular in the torus region) has speed
$\beta \lesssim 0.5$, leading to differences between the comoving and
observer's magnetic field of the order of few percents at most, as can be
seen from Eq.~\ref{eq:bcomov}, well below the level of the quantitative accuracy
with which simulated maps can reproduce observations. 

\section{Emission maps}

In order to show how the presence of small scale fluctuations of the
magnetic field affects the emission pattern of rings and tori, we begin
using a simple toy model analogous to the one used in
\citet{Bucciantini_del-Zanna+05a}, which serves us as a reference for
the case of purely ordered magnetic field. Let us just recall here its
key parameters. The emission is concentrated in a homogeneously
radiating torus around the pulsar. In a spherical reference system
$(r,\theta,\phi)$ centered on the pulsar position, this is equivalent
to the assumption that
\begin{align}
K|B'|^{\alpha+1}=\begin{cases}
 {\rm const}\quad {\rm for}\quad 
(r\sin{\theta}-R_1)^2+(r\cos{\theta})^2 \le R_2^2\\
 0 \quad\quad{\rm otherwise}
\end{cases}
\end{align}
where $R_1$ is the torus principal radius and $R_2$ is the radius of
the cross section (in our model we have adopted the ratio
$R_1/R_2=2.5$). The magnetic field is taken to be purely toroidal, whereas
the flow velocity is constant and purely radial. To take into account
Doppler boosting effects, we
assume a flow speed with $\beta =0.4$, typical of the values inferred
in the torus of Crab nebula and other PWNe
\citep{Del-Zanna_Amato+04a}, and a spectral index $\alpha =0.6$
typical of the brightest X-ray regions in Crab \citep{Mori_Burrows+04a}.

\begin{figure*}
	\centering
	\includegraphics[width=.99\textwidth,bb=0 200 14173 3130,clip]{Pictures/MapToroItot.eps}\\
	\includegraphics[width=.99\textwidth,bb=0 200 14173 3130,clip]{Pictures/MapToroIpol.eps}
	\caption{Upper panels: from left to right the total intensity normalized to its 
          maximum value, for  $\sigma =\sqrt{1/6},\sqrt{1/3},\sqrt{4/3},\sqrt{10/3}$
          (corresponding to ratios of energy in the fluctuating
          components over the one in the ordered toroidal component of
          $0.5,1,4,10$ respectively). Lower panels: the same but for
          the polarized intensity, normalized to the maximum total intensity. 
     }
	\label{fig:toro}
\end{figure*}

In Fig.~\ref{fig:toro} we show the results for various values of the amplitude
of the fluctuating part $\sigma$, in the case of a torus with a
symmetry axis inclined by $30^\circ$ on the plane of the sky [for
reference to a purely ordered case, consider the upper-right panel of Fig.~1 in \citet{Bucciantini_del-Zanna+05a}]. It is immediately evident that for
$\sigma > 1/\sqrt{3}$, corresponding to a case where the fluctuating
components contain the same magnetic energy of the ordered one,  the intensity along the torus changes appreciably with
respect to the ordered case: as the energy in the
fluctuating components rises, the difference in the intensity between
the center of the torus and the sides drops. However, as shown by
\citet{Bucciantini_del-Zanna+05a}, the angular sideways trend of the
luminosity along the torus is also strongly affected by Doppler boosting, so
in principle one can obtain similar trends lowering the flow speed
(see Sect.~\ref{sec:thin}). Interestingly, it
can be shown that for a  flow speed with $\beta \lesssim 0.5$, the two
effects can be disentangled looking also at the brightness difference
between the front and back side of the torus. Such brightness
difference is insensitive to the value of $\sigma$, and depends only
on $\beta$ (see Appendix \ref{sec:app}), so it can be used to set a lower limit to the flow
speed [see again Fig~1 of \citet{Bucciantini_del-Zanna+05a}]. On the other hand the presence of a fluctuating component is
very effective in rising the luminosity at the sides of the torus, thus
the brightness difference between the front and sides can be used to
get another constraint and set limits on $\sigma$. Interestingly, as shown
in  Fig.~\ref{fig:toro}, maps in polarized intensity show little or no variation at all
in their pattern for any value of $\sigma$, what changes is the
polarized intensity (the polarized fraction). This can be easily
understood recalling that in our subgrid model the effect of small scale
fluctuations on the polarized intensity is much smaller than on the
total one. The other important aspect is that,
being a mean-field model, it
does not include possible variances in the polarized properties. So the
polarized direction (the polarized angle) stays unchanged. The ratio of the maximum polarized intensity over
the maximum total intensity goes from $0.7$ for $\sigma=0$ to $0.08$
for $\sigma=\sqrt{10/3}$ ($\delta E=10E$), while the total polarized fraction goes from $38\%$
for $\sigma=0$ to $3\%$ for $\sigma=\sqrt{10/3}$. This shows how
important could be future X-ray polarimetric measures. However some
constraints can be drawn  even now, just using available emission
maps. The known polarized fraction of the
Crab nebula  in X-ray is $PF=19.2\pm1$
\citep{Weisskopf_Silver+78a}. Assuming it is
mostly associated to the emission of the bright torus, this would suggest a
value of $\sigma$ in the range $[0.6-0.8]$, corresponding to a value
of the energy in the fluctuating components of the same order as  the one in
the ordered toroidal field. 

\begin{figure}
	\centering
	\includegraphics[width=.60\textwidth, bb=0 230 700 650, clip]{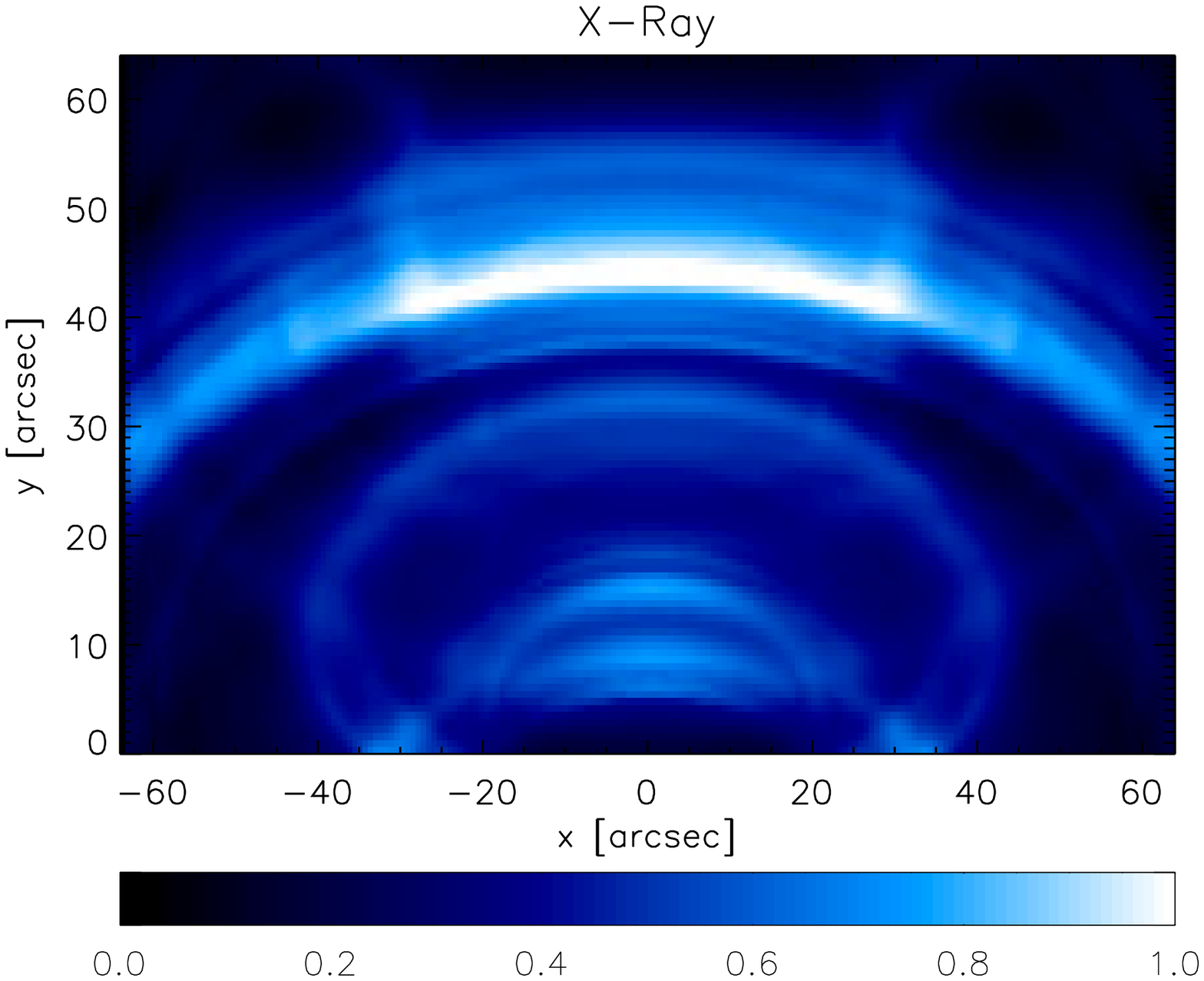}
	\caption{The upper hemisphere of the X-ray surface brightness
          map (1 keV) at $t = 950$ yr, in linear scale, as in Fig 2 of
          \citet{Olmi_Del-Zanna+15a}, but for $\sigma$ raising from
         $0.3 $ at $r=0.6$ ly to $1$ at $r=1.25$ ly. Color-scale normalized to the maximum.
     }
	\label{fig:rmhd}
\end{figure}

As an example of application to more complex multidimensional models
based on relativistic MHD simulations, in Fig.~\ref{fig:rmhd} we show
how the simulated synchrotron map changes due to the inclusion of a
small scale fluctuating magnetic field. The model is the one
described in \citet{Olmi_Del-Zanna+15a}, and targeted to the Crab nebula. For reference to the fully ordered case $(\sigma=0)$
one should take the map shown in Fig.~2 of that same paper. Instead of
just using a
uniform  value of $\sigma$  for the  fluctuating components of the
magnetic field in the
nebula, we have here opted to take a value increasing with distance
from $\sigma=0.3$ at a radius from the center $r=0.6$ ly, to a maximum
value $\sigma=1$ at $r=1.25$ ly. With this choice the inner wisp
region is marginally affected, while fluctuations reach their maximum
in correspondence  with the location of the torus  (see also the next section). The effects of this
small scale turbulent component are twofold: they raise the brightness
of the sides of the various rings and arcs, as already discussed, and  they
also increase the relative brightness of those regions having a
higher value of $\sigma$ (the torus) with respect to the inner ones. 

\section{The ``thin torus'' semi-analytic case}
\label{sec:thin}

An even simpler approach can be attained if we consider the case of a very thin torus, i.e.\ $R_2/R_1\rightarrow0$ (and, as before, constant magnetic field and bulk velocity).
In spite of its simplicity, the thin-torus model maintains most of the features of the more general case, with the advantage of allowing analytical formulae or series expansions that are useful for preliminary surveys of the parameters space.
The points belonging to this torus are defined in the observer's coordinates system by
\begin{equation}
\boldsymbol{r}=\left(R_1\cos\inc\cos\tht+Z\sin\inc,R_1\sin\tht,-R_1\cos\tht\sin\inc+z\cos\inc\right)\,,\label{eq:thinring}
\end{equation}
where $\inc$ is the inclination angle of the symmetry axis with
respect to the plane of the sky, and  where we also allow for a
vertical displacement $z$ of the plane of the torus along the symmetry axis.
The first coordinate is the longitudinal coordinate, positive towards
the observer, while the third one is aligned to  the projection on the sky of the symmetry axis.
If the points in the annulus have velocities with component
$(\bt_Rc,0,\btZ c)$ (in cylindrical coordinates), then the Doppler boosting factor (Eq.~\ref{eq:doppler}) reads
\begin{equation}
D(\tht)=\frac{\sqrt{1-\bt_R^2-\btZ^2}}{1-\bt_R\cos\inc\cos\tht-\btZ\sin\inc}\,,
\end{equation}
while the transverse field in the emitters reference system (Eq.~\ref{eq:bcrossn}) is
\begin{equation}
\mid\boldsymbol{B}'\times\boldsymbol{n}'\mid=B'\sqrt{1-D(\tht)^2\cos\inc^2\sin\tht^2}\,,
\end{equation}
where
\begin{equation}
B'=B\sqrt{1-\bt_R^2-\btZ^2}\,.
\end{equation}
The emissivity, as from Eq.~\ref{eq:j}, then reads
\begin{equation}
j=\left(C\nu^{-\al}B'^{\al+1}\right)\left(1-D(\tht)^2\cos\inc^2\sin\tht^2\right)^{(\al+1)/2}D(\tht)^{\al+2}.
\end{equation}
The power emitted per unit length of the transverse coordinate $y$ ($dI/dy$) can be simply computed as $j(y)\Sigma(y)$, where $\Sigma$ is the cross section parallel to the symmetry axis and to the line of sight
\begin{equation}
\Sigma=(\pi R_2^2)/\sqrt{1-\cos\inc^2\sin\tht^2}\, .
\end{equation}
The above quantities are expressed as explicit functions of $\tht$, but profiles with respect to the variable $y$ are easily obtained by the use of the relation $y=R_1\sin\tht$ in Eq.~\ref{eq:thinring}.

Finally, in order to derive also the other Stokes parameters, we evaluate $q_Y$ and $q_Z$ (Eq.~\ref{eq:stokesang1}) as
\begin{eqnarray}
q_Y&=&B\cos\tht-B(\bt_R\cos\inc+\btZ\sin\inc\cos\tht)\,,	\\
q_Z&=&B\sin\inc\sin\tht-B\,\btZ\sin\tht\,,
\end{eqnarray}
from which, using Eqs.~\ref{eq:stokesang2} the quantities $\cos2\chi$ and $\sin2\chi$ can be derived.

The most relevant aspect to investigate for the present discussion is how the intensity
decreases moving away from the projected axis of symmetry, for the
brighter (front) region of the torus. In Appendix~\ref{sec:app} we
also discuss the intensity ratio between the two regions of the torus
crossing the projected axis (front to back side brightness ratio),
the geometry of the polarization swing
\citep{Bucciantini_del-Zanna+05a} due to the velocities of the
emitters, and other observables in the case of two symmetric rings
(see also Sect.\ref{sec:vela}).

Let us expand the power per unit length to the second order in $y/R_1$, namely
\begin{equation}
\frac{dI(y)}{dy}\simeq\left.\frac{dI(y)}{dy}\right|_{y=0}\left(1+\frac{S}{2}\frac{y^2}{R_1^2}\right)\,.
\end{equation}
The second derivative $S$ is then evaluated as
\begin{equation}
S\!=\!\cos\inc^2\!-\!\frac{(2+\al)\bt_R\cos\inc}{1-\bt_R\cos\inc-\btZ\sin\inc}\!-\!\frac{(1+\al)\cos\inc^2(1-\bt_R^2-\btZ^2)}{(1+\bt_R\cos\inc-\btZ\sin\inc)^2}\,. 
\label{eq:S}
\end{equation}
In other terms, one can define a scale length $y_S=R_1\sqrt{-1/S}$ for
this intensity decrease.

The presence of a random magnetic field component does not affect the
direction of polarization, but contributes both to the behaviour of the polarization fraction and to the pattern of the total intensity, the last one due to the fact that in the presence of a random magnetic field component synchrotron emission is less anisotropic.

\begin{figure}
	\centering
	\includegraphics[width=.42\textwidth]{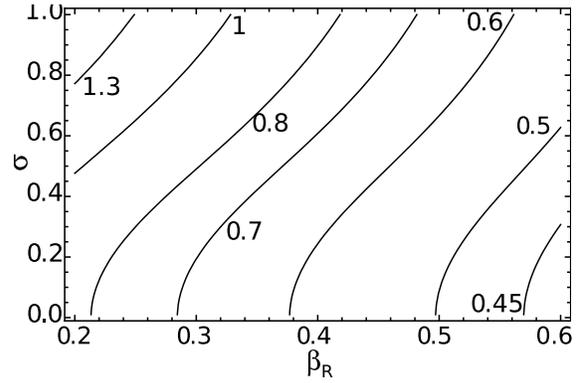}
	\caption{Value of the normalized $y_S/R_1=1/\sqrt{-S}$ brightness scale along
          a thin-ring (with respect to the symmetry axis) given by
          Eq.~\ref{eq:Ss}, as a function of $\bt_R$ and
          $\sigma$. Labels of the contours are the values of
          $y_S/R_1$. The model is computed for $\btZ=0$,
          $\inc=30^\circ$ and $\alpha=0.6$.
     }
	\label{fig:svsv}
\end{figure}

We can derive a generalization of Eq.~\ref{eq:S}, valid for any value of $\sg$
\begin{eqnarray}
S(\sg)&=&\left(\cos\inc^2-\frac{(2+\al)\bt_R\cos\inc}{1-\bt_R\cos\inc-\btZ\sin\inc}\right) +  \nonumber \\	
&-&\frac{(1+\al)\cos\inc^2(1-\bt_R^2-\btZ^2)}{(1+\bt_R\cos\inc-\btZ\sin\inc)^2}\,G_{2,I}(\al,\sg)\,,\label{eq:Ss}
\end{eqnarray}
where
\begin{equation}
G_{2,I}(\al,\sg)=\frac{_1F_1\left(\frac{1-\al}{2},2,-\frac{1}{2\sg^2}\right)}{2\sg^2\,_1F_1\left(-\frac{1+\al}{2},1,-\frac{1}{2\sg^2}\right)}\,.
\end{equation}
$G_{2,I}(\al,0)=1$, so that Eq.~\ref{eq:S} is easily recovered for vanishing $\sg$.
The effect of fluctuations is to decrease the level of anisotropy of
the emission, and therefore to increase the estimated $y_S$ scale: in
this sense, the presence of fluctuations mimics a case with a lower value of $v_R$.
Fig.~\ref{fig:svsv} shows this behaviour.

For small $\sg$ values we propose the following approximation

\begin{eqnarray}
G_{2,I}(\al,\sg)^{-1}&=&1+(1+\al)\sg^2+(1+\al)(1-\al)\sg^4 +  \nonumber \\
&+& 2(1+\al)(1-\al)^2\sg^6 +  \nonumber \\
&+& (1+\al)(1-\al)^2(7-5\al)\sg^8+{\cal O}(\sg^{10})\,.
\end{eqnarray}
Its relative accuracy is, for instance, better than 1\% for $\sg<0.5$, in the range $0.5<\al<2.0$.
In particular it is exact in the case $\al=1$, since $G_{2,I}(1,\sg)=1/(1+2\sg^2)$.

\section{Applications}
\label{sec:vela}

 We apply here our model to the Crab abd Vela
 PWNe. Following an approach similar to the one used by
 \citet{Ng_Romani04a}, we build a simulated synchrotron map to fit the
 three main components seen in X-rays: the torus, the inner ring and
 the jet. In order to obtain a reference image of the Crab nebula as
 shown in Fig.~2, we have aligned and combined 24 Chandra ACIS images
 of different epochs, ranging from 2012 to 2015, as retrieved from the
 Chandra Archive. Since each individual image presents a stripe corresponding to
 the chip gap, as well as the bright line aligned with the saturated
 pulsar image, before adding the images up we have masked these
 critical areas in all of them. Due to the different roll angles of
 the observations,  the stripes in each image show a different
 orientation: therefore we succesfully managed to add them up
 without leaving blind areas. Finally, a pixel-by-pixel correction
 has been applied, to account for the difference in the effective
 exposure time due to the superposition of masked images having
 different orientations and offsets. In this way the medium-large
 scale structure is very well reproduced; only the smaller scales are
 partially washed out, since the nebular structure is highly
 dynamical on those scales. Let us remark here that, given the simplicity of the
 model we have adopted, we have not gone through a full fledged data
 reduction. The data are simply used to derive rough constraints, and
 not to place severe quantitative limits.

\begin{figure*}
	\centering
	\includegraphics[width=.32\textwidth]{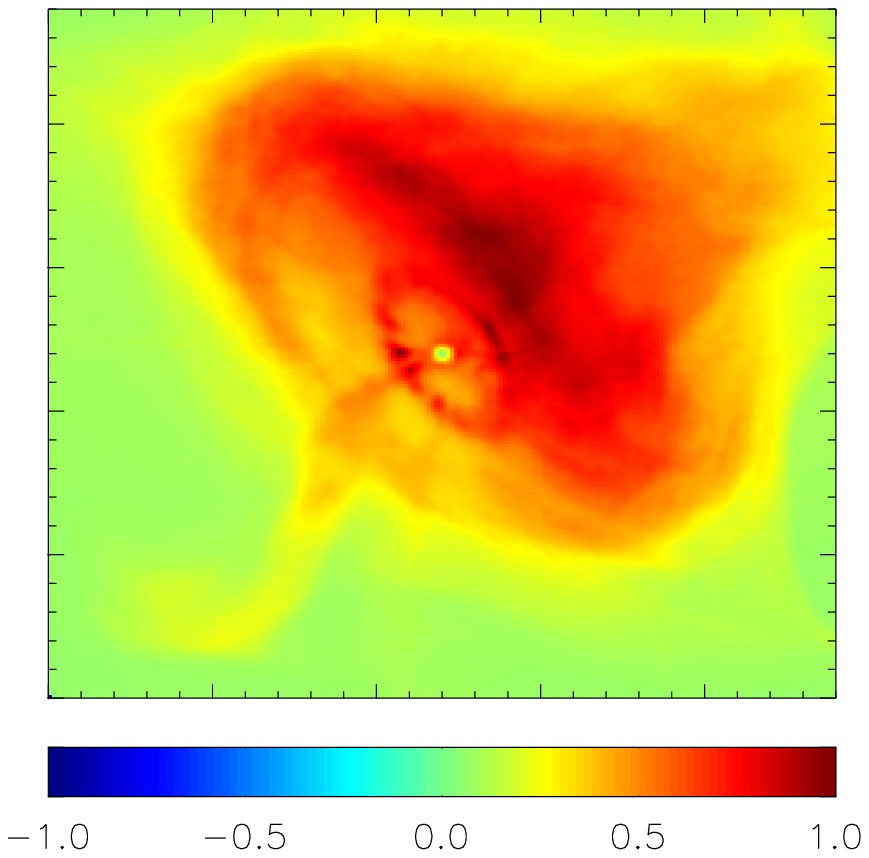}
	\includegraphics[width=.32\textwidth]{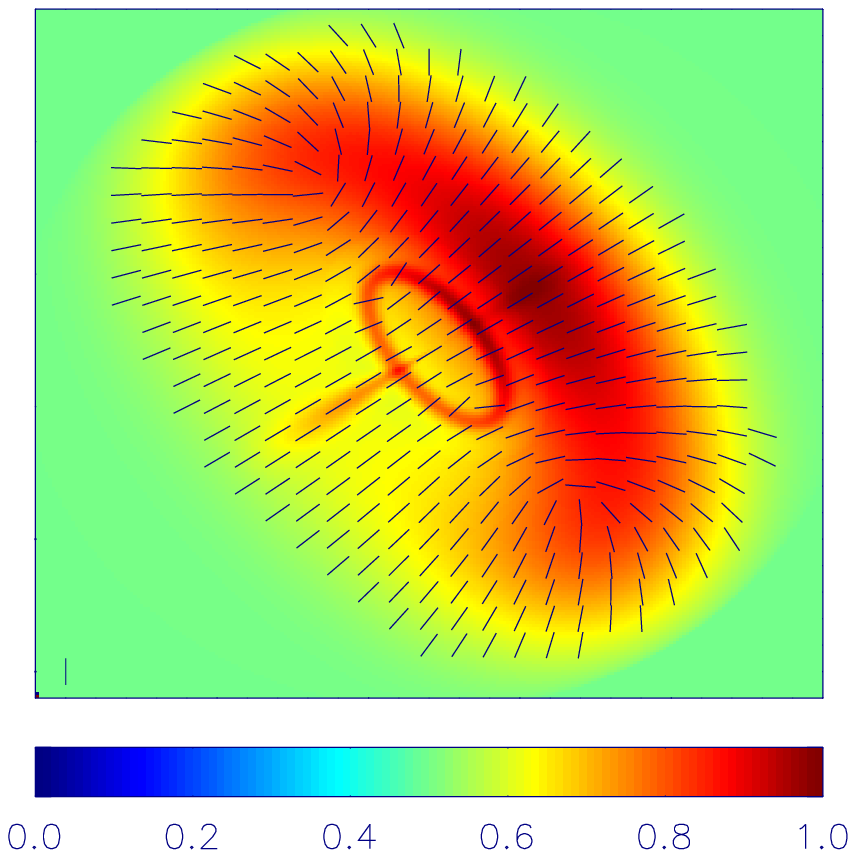}
	\includegraphics[width=.32\textwidth]{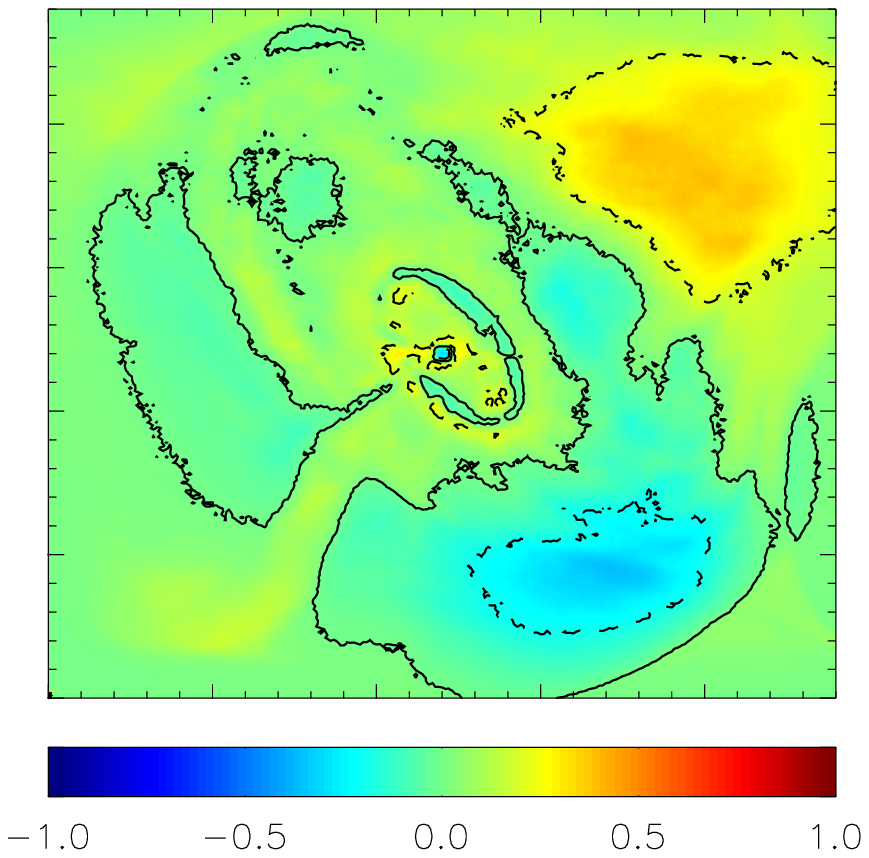}
	\caption{Left panel: CHANDRA X-ray image of the Crab
          nebula. Intensity normalized to the maximum. Middle panel:
          our model of the Crab nebula. In the torus we assume a flow
          velocity of
          $0.35c$, and $\sigma=0.7$. Intensity normalized to the
          maximum, with bars corresponding to the polarized
          direction. Right panel: residuals obtained by subtracting
          the model from  the
          data, normalized to the intensity maximum of the data. The solid
          line represents the zero level. The dashed lines the $\pm 20\%$ level.
     }
	\label{fig:crab}
\end{figure*}

The torus and the inner ring are modeled as discussed in the
 previous section. The only difference is that now  the volume emissivity
 is assumed to fall in a gaussian way, as was
 done by \citet{Ng_Romani04a},  such that

\begin{equation}
K|B'|^{\alpha+1}\propto \exp[-(
(r\sin{\theta}-R_1)^2+(r\cos{\theta})^2)/2R_2^2)] 
\label{eq:exptor}
\end{equation}
The symmetry axis is inclined by $27^\circ$ on the plane of the sky, and
$55^\circ$ with respect to the north \citep{Weisskopf_Hester+00a}. The main torus has $R_1/R_2=3.45$
while for the inner ring we took $R_1/R_2=10$. The same  spectral index $\alpha =0.6$
was used for both \citep{Mori_Burrows+04a}.
  Optical polarization suggests that in the inner ring
 the ordered component of the magnetic field is toroidal. 
There is no information on the structure of the magnetic field in the
torus or jet, even if large scale optical polarization maps are compatible with a toroidal
geometry. This is consistent with simulations (and symmetry arguments) suggesting that the
field should be mostly azimuthal in the torus. On the other hand the jet is
seen to be turbulent and time varying. A model for the X-ray
luminosity of the inner ring was already presented in
\citet{Schweizer_Bucciantini+13a}, where it was shown that it is
possible to reproduce it using a typical  boost speed of $\sim
0.6c\pm 0.1c$. No strong constraints can be placed on the ratio of disordered vs ordered field in the inner
ring, mostly because of the low photon counts and due to the presence
of bright time-varying non axisymmetric features. Thus in the toy model
we set for the inner ring $\sigma= 0.3$ in order to give a polarized
fraction for the ring alone of $\sim 40\%$ as seen in optical (suggesting
that already close to the termination shock about one third of the
magnetic energy is in the small scale fluctuating part). On the other
hand the jet, being a faint feature, can be reproduced with a fully
turbulent field as well as a fully ordered one. Here we decide to
include it (by considering the case of a fully turbulent field) only
to reduce the residuals on the axis.
 Our best fit model for the Crab torus
requires a typical boosting speed of the order of $\simeq0.35c\pm0.05c$ and a
level of fluctuating magnetic field $\sigma \simeq 0.6-0.9$. With
these values we get a total polarized fraction of $\sim (17\pm2) \%$ in
agreement with observations. The value of $\sigma$ can not be raised
above unity otherwise the total polarized fraction becomes smaller
than $15\%$, underestimating the data. 

\begin{figure}
	\centering
	\includegraphics[width=.52\textwidth]{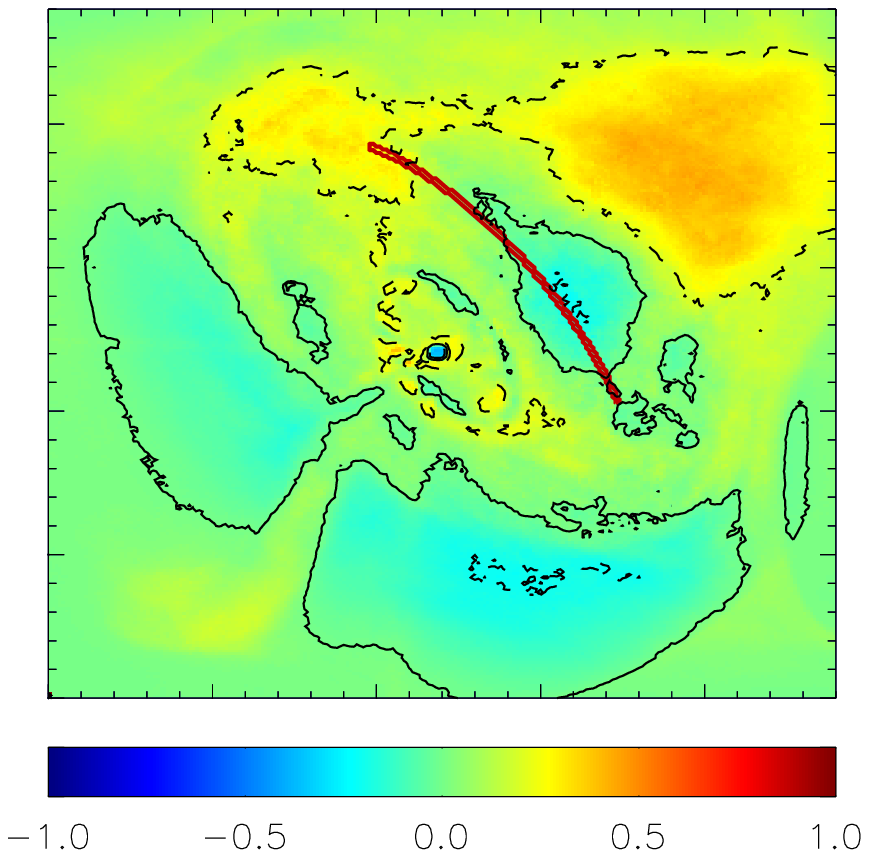}
        \includegraphics[width=.48\textwidth]{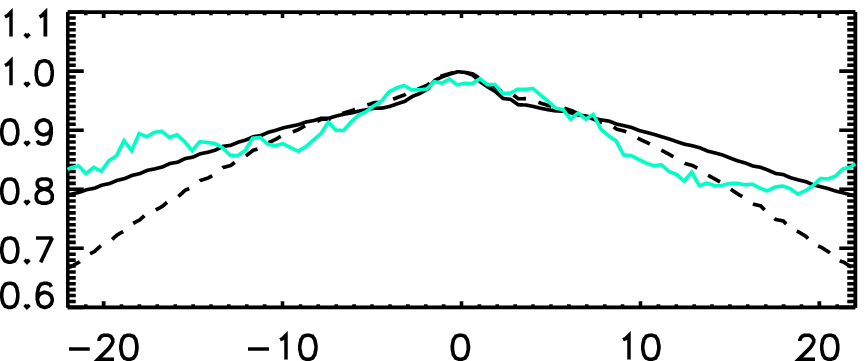}
	\caption{Upper panel: map of the residuals for the Crab nebula
          defined as for
          Fig.~\ref{fig:crab}, but with respect to a model with fully
          ordered toroidal magnetic field. The flow velocity is set to
          $0.3c$ in order to match the front to back  brightens
          difference, while in the torus $\sigma=0$. The solid
          line represents the zero level. The dashed line the $\pm20\%$
          level. Lower panel: comparison of the surface brightness in
          the region defined by the red arc in the upper panel (cyan
          line) to
          the model with $\sigma=0.7$ (solid line) and to the one with
          $\sigma=0$ (dashed line). The curves have been rescaled to
          match the maximum of the intensity. The x-axis is in arcsec
          from the peak. The y-axis is in arbitrary units normalized to
          the peak.
     }
	\label{fig:crab2}
\end{figure}

In Fig.~\ref{fig:crab} we show our
reference image for the Crab nebula, our best match model, and the
residuals  between the data and the model. In Fig.~\ref{fig:crab2} we
show the residuals in the fully ordered case $\sigma=0$, where in order
to get brighter sides of the torus we had to lower the boosting speed
to the possible smallest value $(0.25-0.3)c$. We cannot lower it further. However looking carefully at the residuals, one can
see that even if we are able to get the correct front-to-back side
brightness difference, this model tends to under-predict the front-to-sides
one, on both sides of the torus. 
To make this difference clearer, in the bottom panel of 
Fig.~\ref{fig:crab2} we compare the X-ray surface brightness of  the torus, sampled
along the arc shown in the upper panel of the same figure, to our two models: the
one with a fluctuating small scale field with
$\sigma=0.7$ and the one with a fully ordered field with $\sigma=0$. It is evident that
both models give a reasonable fit in the central (on axis) part of the
torus within $\pm 10$ arcsec from the peak, where they can hardly be distinguished. However the wings of the torus beyond
$\pm 15$ arcsec are slightly undepredicted by the fully ordered case. 
On top of this the integrated polarized fraction for the fully ordered case
is estimated to be $\sim 30\%$. This is much higher than the measured
value of $19\%$ and, given that the torus is by far the brightest feature in
X-ray, it is unlikely that  the low surface brightness diffuse X-ray
emission could provide enough unpolarized radiation to compensate.

As shown in Fig.~\ref{fig:crab}, in the region of the ring and the torus, our best guess model
provides residuals below $15\%$. Given however the presence of a non uniform
and diffused X-ray  nebular emission,  the fact that the torus itself
is brighter on one side, the fact that the ring is not exactly centered
on the pulsar, and that there are non axisymmetric features like the
north-west spur, it is obvious that our axisymmetric model cannot
 provide an accurate fit. But in its simplicity it already indicates
that the brightness profile of the torus points toward a possibly large level of
turbulence (about half of the magnetic energy should be in the small
scale fluctuating part). 
 
\begin{figure}
	\centering
	\includegraphics[width=.52\textwidth]{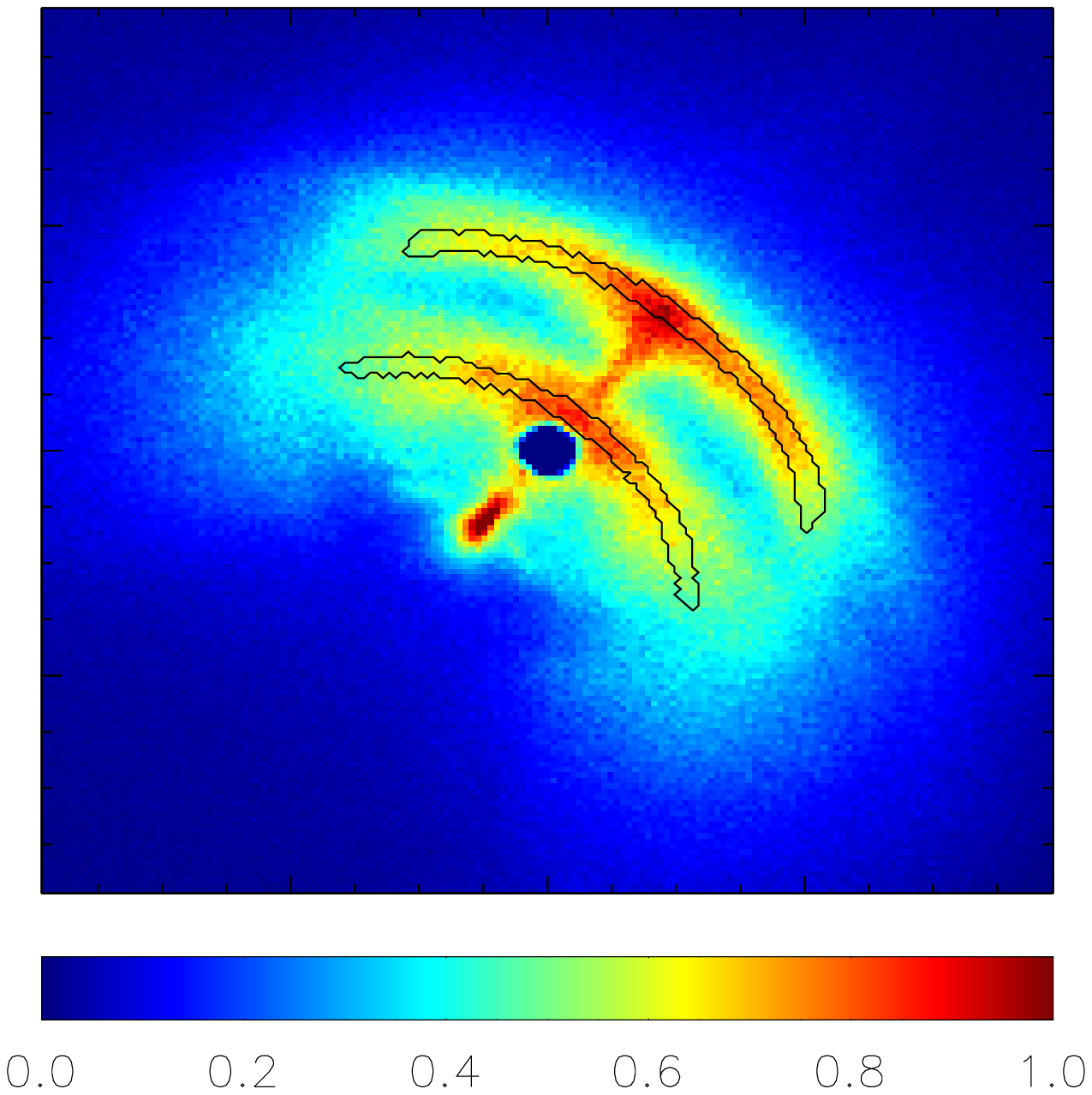}
        \includegraphics[width=.47\textwidth]{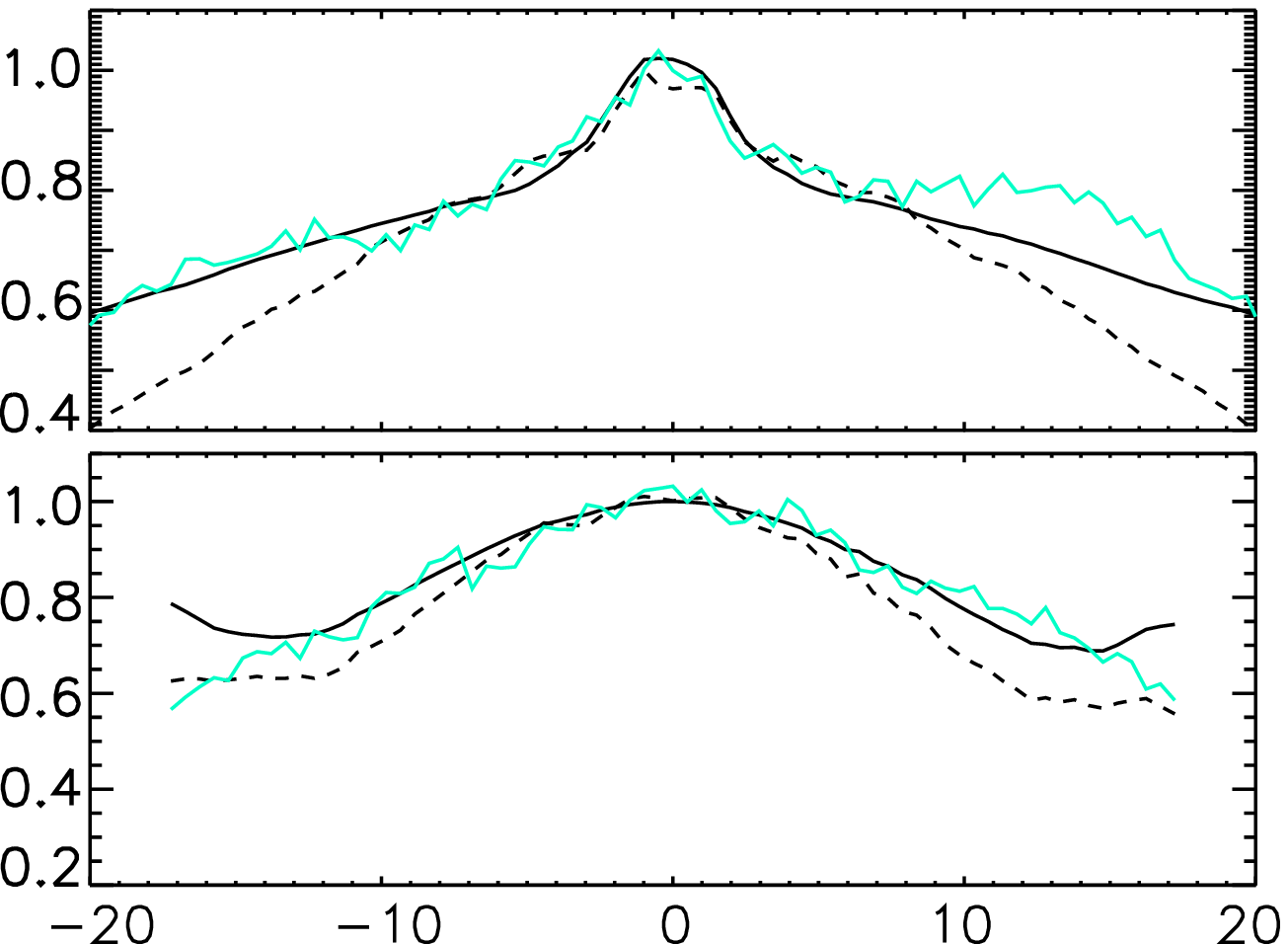}
	\caption{Upper panel: brightness map of the Vela PWN,
          normalized to the maximum. Axes in arcseconds. The central region containing the
          pulsar has been excised. The two arcs represent the region
          of the inner torus and outer torus whence the brightness
          profile was extracted.  Lower panel: comparison of the surface brightness in
          the outer torus (above) and inner torus (below), measured in
          the arcs shown in the upper panel, to
          the model with $\sigma=1.$ (solid line) and to the one with
          $\sigma=0$ (dashed line). The x-axis is in arcsec
          from the peak. The y-axis is in arbitrary units normalized to
          the peak.
     }
	\label{fig:vela}
\end{figure}

We have repeated a similar analysis for the Vela PWN.  To get a
reference image, we have followed a procedure similar to that already described for
the Crab nebula, combining 19 Chandra ACIS images relative to the period from year 2001 to 2010.
Since the relevant areas in individual images now are not affected by
the chip gap, we have simplified the masking procedure with respect to
the previous case. As shown in
Fig.~\ref{fig:vela}, the X-ray nebula is characterized by 
two tori and a small jet. Due to the presence of a large and diffuse X-ray
emission, and to the brightness of the pulsar, we have limited our
investigation to the brightness profile of the two tori, without
attempting a global fit of the emission map. In the upper panel of Fig.~\ref{fig:vela} we
show the regions of the tori where we have extracted the
brightness profiles shown in the bottom panel. We model the tori, as
was done for the Crab nebula, using the gaussian profile of
Eq.~\ref{eq:exptor}, with $R_2/R_1=5.9$ and $5.3$ for the outer and
inner torus respectively. The symmetry axis is
inclined by $33^\circ$ on the plane of the sky and $130^\circ$ with
respect to the north. A jet (and counter-jet) was also introduced
with radial velocity equal to $0.7 c$,  in order to reproduce the
brightness peak observed on axis in the outer torus. The spectral
index is fixed at $\alpha =0.3$ \citep{Kargaltsev_Pavlov04a}. The brightness
difference between the front side (on axis) of the tori and the back
side, constrains the radial flow speed to be higher than $0.35c$.  
In the bottom panel of Fig.~\ref{fig:vela} the brightness profile
of the tori is compared to a fully ordered case $\sigma=0$ with
radial velocity $0.35c$ and to a case with $\sigma=1$ and radial
velocity $\approx 0.5c$.  Again it is evident that the two models begin
to differ substantially beyond $10$ arcsec from the axis. For the
inner torus the difference between the two cases is small. This happens
because the sides of the inner torus are superimposed, along the line
of sight, to back part of the outer torus, so that their brightness does
not decline as fast. On the other hand, the model with a fully ordered
magnetic field under-predicts the brightness of the sides of the outer
torus. A better agreement is achieved by the model with $\sigma=1$. However, we remark here that  the presence of
a large diffuse X-ray emission  does not allow us to perform a
satisfactory global fitting of the emission map using a simple prescription as
Eq.~\ref{eq:exptor}.   We can use our models to provide upper limits to the total integrated
polarized fraction. For $\sigma=0$ we expect $PF \le 23\%$, while
for $\sigma=1$ we find that the polarized fraction should be $\le 6\%$. Only future polarimetric
measures could help us to identify the correct regime.

\section{Conclusions}

Driven by the increasing evidence pointing toward the presence of a
possibly large magnetic turbulence in PWNe, and the interest in future
X-ray polarimetric observations, we have developed here a simple formalism
to simulate the effect of a small scale fluctuating magnetic field on
the emission properties of PWNe, and potentially of other synchrotron
emitting sources, and we have shown how to build emission maps to
be compared with observations. We find that in general there is a
degeneracy between the effects of a turbulent component and the
Doppler boosting. Both regulate how the brightness changes along rings
and tori: the front-to-sides brightness difference can be lowered
either assuming a lower flow speed or a higher level of
turbulence. We showed however that this degeneracy can be partially broken
looking at the front-to-back brightness difference, which depends only
on Doppler boosting. We have applied our analysis to the Crab and
Vela PWNe, showing that models with a sizable fraction of magnetic
energy into a small scale turbulent component seem to provide a
better fit for the tori. In the case of the Crab nebula, where
integrated polarimetric measures are available, our turbulent model
gives a consistent estimate of the polarized fraction. For the Vela
PWN we only provide rough estimates of upper limits for future observations. Our results show that, while
evidence for a turbulent component can already be guessed from
emission maps, future X-ray polarimetric measures, even of just
integrated polarized fraction, will be crucial to set stronger
constraints. This could also help to clarify if the observed
morphological  difference in X-ray  PWNe is possibly related to
different levels of turbulence. 

\section*{Acknowledgements}
The authors acknowledge support from the PRIN-MIUR project prot. 2015L5EE2Y \emph{Multi-scale simulations of high-energy astrophysical plasmas}.

\bibliography{Bib}{}
\bibliographystyle{mn2e}

\appendix
\section{Thin-Ring}
\label{sec:app}

We illustrate here how to derive other observable quantities of
interest for a thin-ring (or a pair of symmetric rings), as a function
of inclination, bulk velocity and level of turbulence.

\begin{figure}
	\centering
	\includegraphics[width=.45\textwidth]{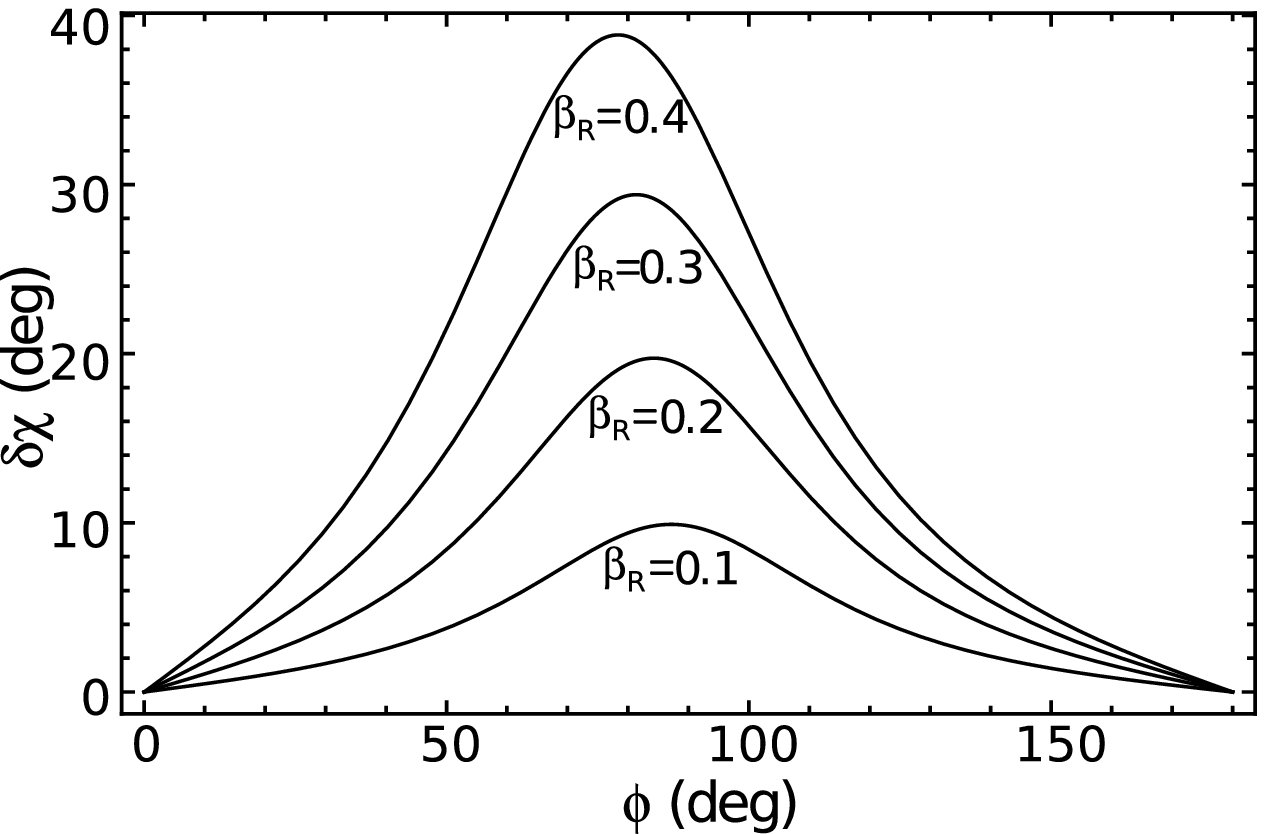}
	\caption{Trend of the polarization angle swing $\dl\chi$ as
          definded in Eq.~\ref{eq:swing}, for different values of the
          velocity $\bt_R$ in the case $\btZ=0$, $\inc=30^\circ$ and
          $\alpha =0.6$.
     }
	\label{fig:swing}
\end{figure}

Following what was done in Sect.~\ref{sec:thin}, we begin with the ratio of the  intensity of the front side ($\tht=0$)
to that on the back side ($\tht=\pi$), which is independent of the value $\sigma$
\begin{equation}
\left.\frac{dI(y)}{dy}\right|_{\tht=0}\bigg/\left.\frac{dI(y)}{dy}\right|_{\tht=\pi}=\left(\frac{1+\bt_R\cos\inc-\btZ\sin\inc}{1-\bt_R\cos\inc-\btZ\sin\inc}\right)^{2+\al}\,.
\end{equation}
Another parameter accessible through polarization measures is the
direction of polarization angle $\chi$ (with respect to the horizontal
axis), that in the absence of motion is simply given by
\begin{equation}
\sin\chi=N\sin\inc\sin\tht\,,\qquad\cos\chi=N\cos\tht\,,
\end{equation}
with a normalization factor such that $N^2=(\sin^2\tht\sin^2\inc+\cos^2\tht)^{-1}$ (the quantity $N$ can either have the positive or the negative sign, corresponding to the fact that the polarization angle is defined modulus $\pi$).
The above formulae simply reflect the geometrical effect of the inclined view.
Instead in the presence of motions this orientation is distorted by
relativistic effects (polarization swing), and the equations give
\begin{eqnarray}
\sin\chi&=&N(\sin\inc-\btZ)\sin\tht\,,	\\
\cos\chi&=&N(\cos\tht-\bt_R\cos\inc-\btZ\sin\inc\cos\tht)\,.
\end{eqnarray}
It is interesting to derive the quantity $\dl\chi$, namely the
deviation of the polarization direction from the purely geometrical
case (zero velocities). Using the above relations, and the formula for the tangent of a sum of angles, one derives
\begin{equation}
\tan(\dl\chi)=\frac{\sin\tht\,\cos\inc\,(\bt_R\sin\inc-\btZ\cos\inc\,\cos\tht)}{(\cos^2\!\tht+\sin^2\!\tht\,\sin^2\!\inc)-(\bt_R\cos\tht\,\cos\inc+\btZ\sin\inc)}\,,
\label{eq:swing}
\end{equation}
with a positive $\dl\chi$ meaning a wider angle, with respect to the
purely geometrical case, from the direction of the mid point of the
brighter front side.
At least in principle, following the behaviour of $\dl\chi$, one could
fit independently $\bt_R$ and $\btZ$ (which could be different from
zero for a ring with a vertical offset from the equator). In
Fig.~\ref{fig:swing} we show this trend.

\begin{figure}
	\centering
	\includegraphics[width=.45\textwidth]{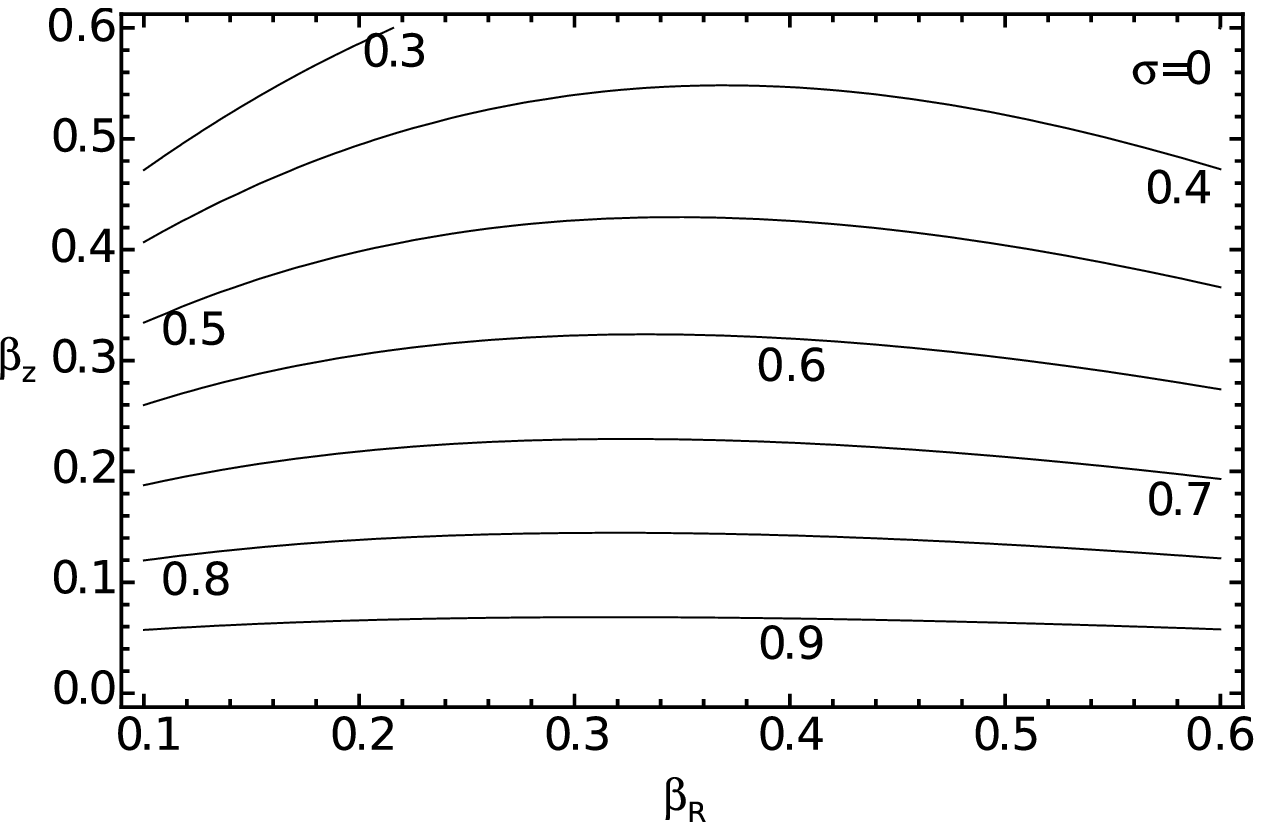}
	\caption{Trend of the $y_{S,+}/y_{S,-}$ ratio for two rings
          symmetric with respect to the equator, as a function of
          $\btZ$ and $\bt_R$, for $\inc=30^\circ$ and $\alpha=0.6$.
     }
	\label{fig:tworing}
\end{figure}

Non vanishing values of $\btZ$  are expected in the case of  a pair of
rings, like for instance in the case of Vela. 
It is then possible to estimate the value of $\btZ$ by comparing the $y_S$ scale length of the two tori.
To keep the system symmetric, let us assume equal values of $\btZ$ in
the two rings, but with opposite signs; in this case,
Fig.~\ref{fig:tworing} shows that the ratio $y_{S,+}/y_{S,-}$ is
strongly dependent on $\btZ$ while it is almost independent of
$\bt_R$, and therefore possibly represents a good diagnostics for $\btZ$.
It can be shown that this quantity is also almost independent of
$\sg$.

Finally, let us discuss the effect of fluctuations on the polarization fraction ($PF$).
By adopting the same kind of procedure as before, namely expanding to the second order in $y$ and then approximating the coefficients in the limit of small values of $\sg$, we get:
\begin{equation}
PF(y)\simeq\frac{\al+1}{\al+5/3}\left(G_{0,PF}(\al,\sg)+\frac{S_P}{2}\frac{y^2}{R_1^2}\right)\,.
\end{equation}
Where
\begin{equation}
G_{0,PF}(\al,\sg)=\frac{(3+\al)\,_1F_1\left(\frac{1-\al}{2},3,-\frac{1}{2\sg^2}\right)}{8\sg^2\,_1F_1\left(-\frac{1-\al}{2},1,-\frac{1}{2\sg^2}\right)}\,,
\end{equation}
whose inverse, for small $\sg$ values, is well approximated by
\begin{eqnarray}
G_{0,PF}(\al,\sg)^{-1}&=&1+2\sg^2+2(1-\al)\sg^4-4\al(1-\al)\sg^6 + 	\nonumber\\
& - & 2(1-\al)^2(3+5\al)\sg^8+{\cal O}(\sg^{10})\,.
\end{eqnarray}
Finally the quantity $S_P$ evaluates
\begin{equation}
S_P(\sg)=-\frac{2\cos^2\!\inc\,(1-\bt_R^2-\btZ^2)}{(1-\bt_R\cos\inc-\btZ\sin\inc)^2}\left(1-G_{2,PF}(\al,\sg)\right)\,,
\end{equation}
where
\begin{eqnarray}
G_{2,PF}(\al,\sg)&=&\frac{(1+\al)\,_1F_1\left(\frac{1-\al}{2},2,-\frac{1}{4\sg^2}\right)}{4\sg^2\,_1F_1\left(-\frac{1+\al}{2},1,-\frac{1}{2\sg^2}\right)} +	\nonumber\\
& +& \frac{(1-\al)\,_1F_1\left(\frac{3-\al}{2},4,-\frac{1}{2\sg^2}\right)}{12\sg^2\,_1F_1\left(\frac{1-\al}{2},3,-\frac{1}{2\sg^2}\right)}\,.
\end{eqnarray}
It can be shown that $G_{2,PF}(\al,\sg)$ tends to 1 for vanishing
$\sg$, namely the polarization fraction is a constant for a completely
ordered field. The power series approximation of this quantity with $\sg$ is
\begin{eqnarray}
G_{2,PF}(\al,\sg)^{-1}&=&1+2\sg^2+4(1-\al)\sg^4 + \nonumber\\
&+& 4(1-\al)(1-3\al)\sg^6 +	\nonumber\\
&-& 8(1-\al)(2+5\al-5\al^2)\sg^8+{\cal O}(\sg^{10}).
\end{eqnarray}
 Clearly this property depends strongly on the assumption of a
 thin-ring. For thick tori (see main text) depolarization effects due
 to integration along the line of sight play instead a more relevant role.

\end{document}